# The mechanical back-action of a spin-wave resonance in a magnetoelastic thin film on a surface acoustic wave

P.G. Gowtham[1], D. Labanowski[1], and S. Salahuddin[1]

[1]Department of Electrical Engineering and Computer Sciences, UC Berkeley

**Abstract**

Surface acoustic waves (SAWs) traveling on the surface of a piezoelectric crystal can, through the magnetoelastic interaction, excite traveling spin-wave resonance in a magnetic film deposited on the substrate. This spin-wave resonance in the magnetic film creates a time dynamic surface stress of magnetoelastic origin that acts back on the surface of the piezoelectric and modifies the SAW propagation. Unlike previous analyses that treat the excitation as a magnon-phonon polariton, here the magnetoelastic film is treated as a perturbation modifying boundary conditions on the SAW. We use acoustical perturbation theory to find closed form expressions for the back-action surface stress and strain fields and the resultant SAW velocity shifts and attenuation. We demonstrate that the shear stress fields associated with this spin-wave back-action also generate effective surface currents on the piezoelectric both in-phase and out-of-phase with the driving SAW potential. Characterization of these surface currents and their applications in determination of the magnetoelastic coupling are discussed. The perturbative calculation is carried out explicitly to first order (a regime corresponding to many experimental situations of current interest) and we provide a sketch of the implications of the theory at higher order.

A considerable amount of interest has developed in harnessing the interaction between gigahertz frequency ultrasound and thin film magnets with appreciable magnetoelastic coupling for various technological applications. Among these applications are the acoustic manipulation and readout of magnetic memory elements[1–4], acoustic driving of magnetic domain walls[5], the acoustic generation of resonant spin-wave excitations[6–14], and magnetic field detectors.[15,16] Some of the interest rests on the point that acoustical wavelengths range in the sub-micron to micron scale at the gigahertz frequencies typical of spin-wave resonance. The coupling of magnetic systems to various classes of lateral mode acoustical resonators (*e.g.,* SAW or contour mode resonators[17,18]) might prove useful in generating various two-dimensional magnetic excitation patterns with sub-micron features. For many of these applications, a clear physical picture and theoretical framework detailing how a magnetic thin film undergoing spin-wave resonance affects the acoustical fields pumping the spin-wave resonance might be important.

In this paper, we calculate this magnetic back-action on the acoustical fields using acoustical perturbation theory. We specifically focus on traveling spin-wave resonance in a magnetoelastic thin film on a piezoelectric substrate excited by surface acoustic waves (SAWs). The acoustical perturbation theory technique and the basic physical picture developed here, however, is applicable to thin magnetic films excited by bulk acoustic waves (BAWs), contour mode resonators, acoustical waveguides, etc. We treat the case of a magnetic thin film of thickness $h$ strained by a SAW of wavelength $\lambda_{SAW}$ traveling on the piezoelectric substrate as shown in Figure 1. We restrict ourselves to situations where $h \ll \lambda_{SAW}$ where the penetration depth of the SAW into the piezoelectric solid is $\sim \lambda_{SAW}$. To the lowest order in the perturbation theory, the strain fields are uniform through the thickness of the film and equal to the SAW strain fields at the surface of the piezoelectric. These strain fields then drive spin-wave resonance in the film. The leading effect that this spin-wave resonance has on the driving elastic field is to generate time-dynamic, thickness-dependent shear and normal stresses of magnetoelastic origin within the film that exert mechanical forces on the top boundary of the piezoelectric. These forces are directly responsible for measured velocity shifts and attenuation of the SAW elastic field.

The physical picture developed here differs from previous analyses of acoustically driven resonance back-action[9,19,20] which treat the spin-wave excitation and SAW as a magnon-phonon polariton propagating in a magnetoelastic semi-infinite solid with magnoelastic coupling $B_{eff}$. An ad-hoc filling factor $F = h/\lambda_{SAW}$ is used to modify the effective coupling of the magnetic to elastic degrees of freedom in the polariton excitation. This is meant to account for the fact that approximately a fraction $F$ of the entire film/substrate volume excited by the SAW is occupied by the magnetoelastic film. It is only this volume fraction that is responsible for SAW attenuation and velocity shifts induced by spin-wave resonance. This method essentially averages over the excited volume of the film/substrate and is reasonable for an estimation of the SAW wave-vector shifts caused by the elastically driven spin-wave resonance. However, it can be shown that the procedure maps to the propagation of a magnon-SAW phonon polariton on the surface of a magnetoelastic semi-infinite solid with weakened magnetoelastic coupling $\sqrt{F}B_{eff}$ and thus neglects the details of the mechanics at the film/substrate interface. We argue that it is *precisely* the back-action stress and strain fields at the film-substrate interface responsible for SAW attenuation and velocity shifts. These fields and their effect on SAW propagation can be calculated using the perturbation formalism without manually inserting a filling factor $F$ or additional fitting parameters (*e.g.,* the ratio of shear to longitudinal strain in the film that might become relevant in thicker films) into the theory.

The formalism for acoustical perturbation theory is developed within the well-established treatises on physical acoustics[21]. Here we go through only the relevant parts of the theory. The starting point is the complex reciprocity relationship for an acoustical wave within a piezoelectric solid

$$\nabla \cdot \left\{ -\mathbf{v}_2^* \cdot \mathbf{T}_1 - \mathbf{v}_1 \cdot \mathbf{T}_2^* + \Phi_2^* (i\omega \mathbf{D}_1) + \Phi_1^* (i\omega \mathbf{D}_2) \right\} = 0 \qquad (1)$$

where all free charges and external forces on the piezoelectric solid are zero and where the electromagnetic quasi-static approximation applies. The quasi-static approximation is justified in SAW experiments as $c/f \gg \lambda_{SAW}$ where $c$ is the speed of light. We then define $\mathbf{v}_1 \to \mathbf{v}$, $\mathbf{T}_1 \to \mathbf{T}$ as the particle velocity field, stress tensor, etc. arising from solid deformations of the unperturbed SAW propagating on the surface of the piezoelectric substrate (i.e. the velocity field of the SAW substrate without the magnetic film on top). The fields $\mathbf{v}_2 \to \mathbf{v}'$, $\mathbf{T}_2 \to \mathbf{T}'$, etc. are the perturbed fields within the piezoelectric substrate due to the presence of the magnetoelastic film at the surface. The complex reciprocity theorem holds between the two field solutions and their source terms (in this case source terms are zero) provided that the piezoelectric substrate is considered lossless. The reciprocity is correct even in the presence of a lossy perturbation at the surface. The perturbed and unperturbed velocity field, stress fields, etc of a SAW traveling in the $z$ direction are:

$$\begin{aligned} \mathbf{v} &= \mathbf{v}(y) e^{-i\beta z + i\omega t} \\ \mathbf{v}' &= \mathbf{v}'(y) e^{-i\beta' z + i\omega t} \\ &\ldots \end{aligned} \qquad (2)$$

In order to find the shift in the wavenumber $\beta$, Eqn. (1) is integrated over the thickness of the piezoelectric substrate and we have assumed that there is no $x$ dependence in the fields as appropriate for plane-wave SAW propagation. It follows from Eqns. (1) and (2) that the wave-vector shift can be expressed as

$$\Delta\beta = \beta' - \beta = -i \frac{\left\{-\mathbf{v}^*\cdot\mathbf{T}' - \mathbf{v}'\cdot\mathbf{T}^* + \Phi^*(i\omega\mathbf{D}') + \Phi'(i\omega\mathbf{D})^*\right\}\cdot\hat{\mathbf{y}}\Big|_0^b}{\int_0^b \left\{-\mathbf{v}^*\cdot\mathbf{T}' - \mathbf{v}'\cdot\mathbf{T}^* + \Phi^*(i\omega\mathbf{D}') + \Phi'(i\omega\mathbf{D})^*\right\}\cdot\hat{\mathbf{z}}\,dy} \quad (3)$$

Given that the perturbed solutions are assumed to be close to the unperturbed solutions, it is reasonable to set the perturbed field equal to the unperturbed fields in the denominator. The denominator then becomes $2\int_0^b \left\{-\mathbf{v}^*\cdot\mathbf{T} + \Phi(i\omega\mathbf{D})^*\right\}\cdot\hat{\mathbf{z}}\,dy = 4P_{SAW}$ where $P_{SAW}$ is the power flow of the SAW. The numerator is additive in the contributions from mechanical and electrical components. We are only considering contributions from the elastic and magnetoelastic part of the dynamics and thus exclude the wavenumber shifts in Eqn. (3) arising from the electric displacement field and charge dynamics on the surface of the ferromagnetic film that would be present if the magnetic film were conductive. For SAW propagation, we need only concern ourselves with the top surface at y = 0 where the perturbing film is situated. Eqn. (3) then reduces to

$$\Delta\beta = \beta' - \beta = -i \frac{\mathbf{v}^*\cdot\mathbf{T}'\cdot\hat{\mathbf{y}}\big|_{y=0}}{4P_{SAW}} \quad (4)$$

The shifts in the wavenumber of the SAW is thus directly related to stress at the surface of the piezoelectric caused by the perturbing magnetoelastic thin film (the unperturbed traction force $\mathbf{T}\cdot\hat{\mathbf{y}}\big|_{y=0} = 0$ due to stress-free boundary conditions at the substrate surface).

A calculation of the traction force acting at the interface requires a solution to the stress fields within the magnetoelastic film. We express these stress fields in terms of the unperturbed particle velocities of the SAW at $y = 0$ and solve for $\Delta\beta$ to lowest order. The first field equation governing dynamics in the magnetoelastic thin film is $\rho \frac{d\mathbf{v}'}{dt} = \nabla \cdot \mathbf{T}'$. Component-wise this yields:

$$i\omega\rho v_x' = \frac{\partial}{\partial y} T_{yx}' - i\beta T_{zx}'$$
$$i\omega\rho v_y' = \frac{\partial}{\partial y} T_{yy}' - i\beta T_{zy}' \quad (5)$$
$$i\omega\rho v_z' = \frac{\partial}{\partial y} T_{yz}' - i\beta T_{zz}'$$

The second set of field equations define the stress tensor $T_{ij}' = \left(\frac{\partial F}{\partial \varepsilon_{ij}}\right)_T$ in the magnetoelastic thin film, where $\varepsilon_{ij}$ is the mechanical strain tensor and $F$ is the free energy of the magnetoelastic solid, is:

$$F = E - TS = \sigma_{ij}^{mech'} \varepsilon_{ij}' + B_{ij} m_i m_j \varepsilon_{ij}' + K_u m_z^2 - M_s m_i H_i^{app} + \left(2\pi M_s^2 - K_\perp\right) m_y^2 \quad (6)$$

The stress $\sigma_{ij}^{mech} = c_{ijkl}\varepsilon_{kl}$ is the mechanical stress generated by linear elasticity, $B_{ij}$ is the magnetoelastic coupling, $K_u$ is the in-plane anisotropy energy density, $M_s$ is the saturation magnetization, $m_i$ and $H_i^{app}$ are the components of the applied field and components of the magnetization normalized to the unit sphere respectively, and $K_\perp$ is the perpendicular anisotropy energy density. We assume for the remainder of the paper that the film is in-plane magnetized with $K_\perp < 2\pi M_s^2$ and $H_{app}$ in the film plane, and that $K_u > 0$ implying that $x$ is the film's magnetic easy axis. The total stress tensor is $T_{ij}' = c_{ijkl}'\varepsilon_{kl}' + B_{ij}'m_i m_j$. An inversion of this equation to solve for $\varepsilon_{kl}$ and using the relation $\dfrac{\partial \varepsilon}{\partial t} = \nabla \mathbf{v}$ between the strain and particle velocity fields yields the second set of field equations

$$\frac{1}{i\omega}\left(\frac{\partial v_l}{\partial x_k}\right) = s_{klij}T_{ij} - s_{klij}B_{ij}m_i m_j \qquad (7)$$

where $s_{klij}$ are elements of the elastic compliance tensor. If the magnetoelastic part of the free energy $F_{ME}$ has symmetry in the $xz$ plane and the shear terms are all governed by the same coupling term $B_2$, then $F_{ME}$ reduces to:

$$F_{ME} = B_{11}m_x^2\varepsilon_{xx} + B_{11}m_z^2\varepsilon zz + B_{12}m_y^2\varepsilon_{yy} + B_2 m_y m_z \varepsilon_{yz} + B_2 m_x m_y \varepsilon_{yx} + B_2 m_x m_z \varepsilon_{zx}. \qquad (8)$$

Since none of the field quantities can have a dependence on $x$, we use $\varepsilon_{xx} = 0$ to eliminate $T_{xx} - B_{11}m_x^2$ from the remaining equations. The resulting component-wise expression for the second set of field equations is:

$$-\left(\frac{\beta'}{\omega}\right)v'_z = \left(\frac{s'_{12}s'_{11}-s'^{2}_{12}}{s'_{11}}\right)(T'_{yy}-B_{12}m_y^2)+\left(\frac{s'^{2}_{11}-s'^{2}_{12}}{s'_{11}}\right)(T'_{zz}-B_{11}m_z^2)$$

$$-\left(\frac{\beta'}{\omega}\right)v'_x = s'_{44}(T'_{zx}-B_2 m_x m_z)$$

$$-\left(\frac{\beta'}{\omega}\right)v'_y = s'_{44}(T'_{zy}-B_2 m_y m_z)$$

$$\left(\frac{1}{i\omega}\right)\frac{\partial v'_y}{\partial y} = \left(\frac{s'^{2}_{11}-s'^{2}_{12}}{s'_{11}}\right)(T'_{yy}-B_{12}m_y^2)+\left(\frac{s'_{12}s'_{11}-s'^{2}_{12}}{s'_{11}}\right)(T'_{zz}-B_{11}m_z^2) \quad (9)$$

$$\left(\frac{1}{i\omega}\right)\frac{\partial v'_x}{\partial y} = s'_{44}(T'_{yx}-B_2 m_x m_y)$$

$$\left(\frac{1}{i\omega}\right)\frac{\partial v'_z}{\partial y} = s'_{44}(T'_{yz}-B_2 m_y m_z)$$

We now use Eqns. (5) and (9) to solve for the stress fields to first order and expand the velocity and stress tensor fields in a power series in film thickness:

$$\mathbf{v}' = \mathbf{v}'^{(0)} + \mathbf{v}'^{(1)}(y+h) + \mathbf{v}'^{(2)}(y+h)^2 + ...$$
$$\mathbf{T}' = \mathbf{T}'^{(0)} + \mathbf{T}'^{(1)}(y+h) + \mathbf{T}'^{(2)}(y+h)^2 + ... \quad (10)$$

The stress tensor at $y=-h$ (the top surface of the film) is such that $\mathbf{T}'\cdot\hat{\mathbf{y}}\big|_{y=-h}=0$ due to stress-free boundary conditions implying the zeroth order contribution to the traction force at the film/substrate interface $\mathbf{T}'\cdot\hat{\mathbf{y}}\big|_{y=0}=0$. We thus solve for the stress tensor $\mathbf{T}'^{(1)}\cdot\hat{\mathbf{y}}$ at first order. Eqn. (5) and the first two formulas in Eqn. (10) provide the sufficient equations that can be used to solve for the three components of $\mathbf{T}'^{(1)}\cdot\hat{\mathbf{y}}$. We drop the term of the magnetoelastic stress going as $B_{12}m_y^2$ given that we are restricting ourselves to in-plane magnetized films. The equations then become:

$$-\left(\frac{\beta'}{\omega}\right)v_x'^{(0)} = s_{44}'T_{zx}'^{(0)} - s_{44}'B_2 m_z m_x$$

$$-\left(\frac{\beta'}{\omega}\right)v_z'^{(0)} = \left(\frac{s_{11}'^2 - s_{12}'^2}{s_{11}'}\right)T_{zz}'^{(0)} - \left(\frac{s_{11}'^2 - s_{12}'^2}{s_{11}'}\right)B_{11}m_z^2 \quad (11)$$

$$i\omega\rho v_x'^{(0)} = T_{yx}'^{(1)} - i\beta'T_{zx}'^{(0)}$$

$$i\omega\rho v_y'^{(0)} = T_{yy}'^{(1)}$$

$$i\omega\rho v_z'^{(0)} = T_{yz}'^{(1)} - i\beta'T_{zz}'^{(0)}$$

The term $T_{zy}'^{(0)} = 0$ vanishes as $\mathbf{T}^{(0)} \cdot \hat{\mathbf{y}}\big|_{y=0} = 0$ (i.e. at lowest order surface shear stress vanishes) and the stress tensor is symmetric. The components of the magnetoelastic stresses $s_{44}'B_2 m_z m_x$ and $\left(\frac{s_{11}'^2 - s_{12}'^2}{s_{11}'}\right)B_{11}m_z^2$ of Eqn. (11) causing back-action traction forces on the SAW at order $\mathbf{T}^{(1)}$ are $s_{44}'B_2\left(m_z^{(0)}\delta m_x + m_x^{(0)}\delta m_z\right)$ and $\left(\frac{s_{11}'^2 - s_{12}'^2}{s_{11}'}\right)2B_{11}m_z^{(0)}\delta m_z$ respectively. We define $m_x^{(0)}$ and $\delta m_x$ ($m_z^{(0)}$ and $\delta m_z$) as the $x$ ($z$) component of the in-plane equilibrium magnetization and the excited spin-wave amplitude respectively. The $xz$ magnetoelastic shear back-action stresses are present at lowest order for shear horizontal SAWs (SH-SAWs) and Love waves. Here we focus on the Rayleigh SAW for the sake of comparison with previous studies. The Rayleigh SAW contains particle velocity components in the $y$ and $z$ directions only ($v_x^{(0)} = 0$). Rearranging of terms in Eqn. (11) yields expressions for the first order stress tensor causing the perturbing surface traction on the SAW:

$$T'^{(1)}_{yz} = i\omega\left[\rho' - \frac{s'_{11}}{s'^2_{11} - s'^2_{12}} \cdot \frac{1}{V^2_{SAW}}\right]v'^{(0)}_z + i2\beta' B_{11} m^{(0)}_z \delta m_z$$

$$T'^{(1)}_{yy} = i\omega\rho' v'^{(0)}_y \tag{12}$$

$$T'^{(1)}_{yx} = \left[i\omega\rho' - \frac{1}{s'_{44}} \cdot \frac{1}{V^2_{SAW}}\right]v'^{(0)}_x + i\beta' B_2\left(m^{(0)}_z \delta m_x + m^{(0)}_x \delta m_z\right) = 0$$

The magnetoelastic terms in the expression for $T'^{(1)}_{yx}$ are ignored as they can be shown to be proportional to $v^{(0)}_x$. The spin-wave amplitude $\delta m_z$ is excited by an internal RF field arising from the dynamic strain in the film caused by the Rayleigh SAW propagating on the piezoelectric substrate. To lowest order in the perturbation theory, this internal effective magnetoelastic pump field can be expressed in terms of the unperturbed particle velocity field present at the surface of the piezoelectric substrate. The driven spin-wave amplitude can then be solved for in terms of these unperturbed SAW velocity fields. This is accomplished by a linearization of the Landau-Lifshitz-Gilbert (LLG) equations for spin-wave dynamics about the equilibrium magnetization $\mathbf{m_0}$

$$\frac{d\delta\mathbf{m}(\mathbf{r})}{dt} = -\gamma\delta\mathbf{m}(\mathbf{r})\times\mathbf{H}_{eff}(\mathbf{r}) + \Gamma(\boldsymbol{\beta},\mathbf{m_0})\delta\mathbf{m}(\mathbf{r})\times\frac{d\delta\mathbf{m}(\mathbf{r})}{dt}, \tag{13}$$

where $\gamma$ is the effective gyromagetic ratio (for the remainder of the paper taken to be the free electron value appropriate to metallic transition ferromagnets), $\Gamma(\boldsymbol{\beta},\mathbf{m_0})$ is the spin-wave damping at propagation vector $\boldsymbol{\beta}$ at equilibrium magnetization orientation $\mathbf{m_0}$, and where $\mathbf{H}_{eff}(\mathbf{r})$ is the spatially varying effective magnetic field acting on the magnetization. The effective magnetic field contains terms arising from the applied field, internal anisotropy fields, the magnetoelastic interaction, and leading order spin-spin interactions (i.e., dipolar field and exchange contributions). We define a new coordinate system $\eta\zeta\xi$ specified in Figure 2 where $\mathbf{m}_0$ lies along $\xi$ making an angle $\varphi_0$ with respect to the $z$ axis. The $\zeta$ axis is out of the film plane and $\eta$ defines the axis orthogonal to $\mathbf{m}_0$ in the film-plane. The Rayleigh SAW creates a time-varying effective magnetic pump field

$$\mathbf{h}_{rf}(\mathbf{r},t) = -(\partial F_{ME}/\partial \mathbf{m})/M_s = \left(\frac{\beta}{\omega}\right)\left(\frac{2B_{11}}{M_s}\right)v_z^{(0)}\sin\phi_0\cos\phi_0 e^{-i\beta z+i\omega t}\hat{\boldsymbol{\eta}} \quad (14)$$

where the only term in $F_{ME}$ that is non-zero in the in-plane magnetized case goes as $B_{11}m_z^2\varepsilon_{zz}$. The final form of Eqn. (14) has been derived in other work[6,19] and the only difference is that we have substituted the unperturbed strain field $\varepsilon_{zz}^{(0)}$ for the unperturbed particle velocity field $v_z^{(0)}$ at the substrate surface. The pump field then drives a spin-wave resonance $\delta\mathbf{m}(z,t) = \delta\mathbf{m}e^{-i\beta z+i\omega t}$ where $\delta\mathbf{m} = \delta m_\eta\hat{\boldsymbol{\eta}} + \delta m_\zeta\hat{\boldsymbol{\zeta}}$. A solution of the components of the spin-wave amplitude requires a self-consistent solution of the LLG equation along with the magnetostatic equations for long-range dipolar fields. We point out that the amplitude $\delta\mathbf{m}$ is a thickness-averaged spin-wave amplitude. Even under the influence of a magnetoelastic pump field $\mathbf{h}_{rf}(\mathbf{r},t)$ that is uniform in $y$, boundary conditions on magnetostatic potentials and considerations of surface spin pinning will create a spin-wave amplitude profile in the $y$ direction going as $\beta h \ll 1$.[22] Effects of this $y$ magnetization profile will show up directly in the stress tensor at second order. The relationship between these thickness-averaged spin-wave amplitudes and the driving magnetoelastic pump field is given by the Polder susceptibility $\chi$:

$$\begin{pmatrix}\delta m_\eta \\ \delta m_\zeta\end{pmatrix} = \left\{\begin{pmatrix}\chi'_{\eta\eta} & \chi'_{\eta\zeta} \\ \chi'_{\zeta\eta} & \chi'_{\zeta\zeta}\end{pmatrix} + i\begin{pmatrix}\chi''_{\eta\eta} & \chi''_{\eta\zeta} \\ \chi''_{\zeta\eta} & \chi''_{\zeta\zeta}\end{pmatrix}\right\}\begin{pmatrix}h_{RF} \\ 0\end{pmatrix}, \quad (15)$$

The susceptibility components $\chi'$ and $\chi''$ are the relevant real and imaginary components of the traveling spin-wave susceptibility excited about equilibrium $\mathbf{m}_0$ with propagation vector $\boldsymbol{\beta}$. We emphasize that the *y* dependence of the spin-wave amplitudes, while not directly affecting stress fields at first order, will create appreciable modifications to $\chi$ and thus impact $\mathbf{T}^{(1)}$ through $\chi$. The only component of the traveling spin-wave that contributes to $\mathbf{T}'^{(1)} \cdot \hat{\mathbf{y}}$ in Eqn. (12) is $\delta m_z = -\delta m_\eta \sin\phi_0$ in the stress tensor component $T'^{(1)}_{yz}$. It can be shown that Eqns. (12), (14), and (15) yield an expression for $T'^{(1)}_{yz}$ in terms of the unperturbed SAW particle velocity field:

$$T'^{(1)}_{yz} = i\omega \left\{ \left[ \rho' - \frac{s'_{11}}{s'^2_{11} - s'^2_{12}} \cdot \frac{1}{V^2_{SAW}} \right] - \frac{4B^2_{11}}{M_s} \cdot \frac{1}{V^2_{SAW}} \sin^2\phi_0 \cos^2\phi_0 \left[ \chi'_{\eta\eta} + i\chi''_{\eta\eta} \right] \right\} v'^{(0)}_z \quad (16)$$

where $V_{SAW} = \beta/\omega$ and $m^{(0)}_z = \cos\phi_0$. The real part $\chi'_{\eta\eta}$ and imaginary part $\chi''_{\eta\eta}$ can be shown to be:

$$\chi'_{\eta\eta} = \gamma^2 \frac{\left[ \omega^2_{res} - \omega^2 \right]\Upsilon + \left( \Gamma(\boldsymbol{\beta}, \mathbf{m}_0)\omega \right)^2 (\Psi + \Upsilon)}{\left[ \omega_{res}^2 - \omega^2 \right]^2 + \left[ \Gamma(\boldsymbol{\beta}, \mathbf{m}_0)\omega\gamma(\Psi + \Upsilon) \right]^2}$$

$$\chi''_{\eta\eta} = \gamma \frac{\omega\Gamma(\mathbf{q}_{\parallel}, \mathbf{m}_0)\left( \gamma^2\Upsilon^2 + \omega^2 \right)}{\left[ \omega_{res}^2 - \omega^2 \right]^2 + \left[ \Gamma(\boldsymbol{\beta}, \mathbf{m}_0)\omega\gamma(\Psi + \Upsilon) \right]^2} \quad (17)$$

The traveling spin-wave resonance frequency $\omega_{res} = \gamma\sqrt{\Psi\Upsilon}$ and the quantities $\Psi$ and $\Upsilon$ are:

$$\Psi = -H_k \cos(2\phi_0) + H_{app}\cos(\phi_0 - \phi_H) + \frac{2A_{ex}}{M_s}\beta^2 + 2\pi M_s \beta h \sin^2\phi_0$$

$$\Upsilon = H_k \sin^2\phi_0 + H_{app}\cos(\phi_0 - \phi_H) + \frac{2A_{ex}}{M_s}\beta^2 + \left( 4\pi M_s - \frac{2K_\perp}{M_s} \right) - 4\pi M_s \left( \frac{\beta h}{2} \right). \quad (18)$$

where $A_{ex}$ is the magnetic exchange stiffness, $H_k = 2K_u/M_s$ is the in-plane anisotropy field with the in-plane anisotropy along the $x$ axis. The quantities $2\pi M_s \beta h \sin^2\phi_0$ and $4\pi M_s \left(\dfrac{\beta h}{2}\right)$ in $\Psi$ and $\Upsilon$ are terms of long-range dipolar origin and arise due to the $y$ dependence of the spin-wave profile. In the mid sub-micron (500 nm) to micron regime, these quantities create appreciable corrections to $\chi$.[11]

The wavenumber shift $\Delta\beta$ in Eqn. (4) is then:

$$\Delta\beta = -i\frac{\mathbf{v}^*\cdot\mathbf{T}'\cdot\hat{\mathbf{y}}\big|_{y=0}}{4P_{SAW}} = \frac{\omega h}{4P_{SAW}}\left\{\begin{array}{l}\rho'\left|v_y^{(0)}\right|^2_{y=0} + \left(\rho' - \dfrac{s'_{11}}{s'^2_{11}-s'^2_{12}}\cdot\dfrac{1}{V^2_{SAW}}\right)\left|v_z^{(0)}\right|^2_{y=0} \\ -\dfrac{4B^2_{11}}{M_s}\cdot\dfrac{1}{V^2_{SAW}}\sin^2\phi_0\cos^2\phi_0\left[\chi'_{\eta\eta}+i\chi''_{\eta\eta}\right]\left|v_z^{(0)}\right|^2_{y=0}\end{array}\right\} \quad (19)$$

The quantities $\left|v_y^{(0)}\right|^2_{y=0}/P_{SAW}$, and $\left|v_z^{(0)}\right|^2_{y=0}/P_{SAW}$ can be expressed analytically as $|c_y|^2\omega$ and $|c_z|^2\omega$ where $|c_y|^2$ and $|c_z|^2$ have units of $[\text{g/cm}]^{-1}$ and depend on the electromechanical properties of the substrate. Values for $V_{SAW}$, $|c_x|^2$, $|c_y|^2$ and $|c_z|^2$ for some SAW substrates are provided in Table 1.

| SAW Substrate and Cut (Propagation Direction) | $V_{SAW}$ ($10^5$ cm/s) | $|c_x|^2$ ($10^{-13}$ cm/g) | $|c_y|^2$ ($10^{-13}$ cm/g) | $|c_z|^2$ ($10^{-13}$ cm/g) |
|---|---|---|---|---|
| YZ-Cut LiNbO$_3$, Z-prop | 3.488 | 0 | 6.891 | 3.158 |
| [001]-cut Bi$_{12}$GeO$_{20}$, [110]-prop | 1.680 | 0 | 17.331 | 6.436 |
| Y-cut Quartz, X-prop | 3.158 | 2.062 | 18.809 | 8.422 |

Table 1. Propagation characteristics for a few SAW substrates/cuts and propagation directions. Adapted from Auld.[21]

The first two terms in Eqn. (19) are shifts due to the standard mass loading of the SAW by a lossless isotropic thin film of thickness $h$ given a certain mass density $\rho'$ and compliance tensor $\mathbf{s}'$. The last term is due to the mechanical back-action of the elastically driven traveling spin-wave resonance on the Rayleigh SAW and we isolate it from the rest of Eqn. (19):

$$\Delta \beta_{ME}^{Rayleigh} = -\omega^2 h \frac{B_{11}^2}{M_s} \cdot \frac{1}{V_{SAW}^2} \sin^2 \phi_0 \cos^2 \phi_0 \left[ \chi'_{\eta\eta} + i\chi''_{\eta\eta} \right] |c_z|^2 = Z_{ME}^{Rayleigh} |c_z|^2 \omega \qquad (20)$$

$$Z_{ME}^{Rayleigh} = -\omega h \frac{B_{11}^2}{M_s} \cdot \frac{1}{V_{SAW}^2} \sin^2 \phi_0 \cos^2 \phi_0 \left[ \chi'_{\eta\eta} + i\chi''_{\eta\eta} \right] \qquad (21)$$

where $Z_{ME}^{Rayleigh}$ is the SAW electromechanical transmission line impedance due to spin-wave backaction. The wavenumber shift $\Delta \beta_{ME}^{Rayleigh}$ is complex and thus the elastic excitation of the traveling spin-wave resonance modifies the velocity of the SAW and cause an exponential attenuation. The attenuation of the SAW has a rather simple physical interpretation. The SAW, under the right external field conditions, drives a spin-wave resonance via the magnetoelastic interaction in the magnetic film. Part of this response will be out of phase with the SAW elastic drive field due to the spin-wave damping. Thus a thickness-dependent and time-varying $yz$ magnetoelastic shear stress develops in the film and generates a back-action traction force at the surface of the piezoelectric out of phase with the driving Rayleigh SAW field. This out-of-phase traction force dampens the SAW. This also implies, through Eqn. (20), an electromechanical transmission line current that is generated out of phase with the SAW surface potential $\Phi$ due to the spin-wave back action. The power of the Rayleigh SAW per unit width attenuates under the magnetic film due to magnetoelastic back-action as

$$P(z) = P_{SAW} \exp[2\,\text{Im}(\Delta\beta_{ME}^{Rayleigh})z]$$
$$= P_{SAW} \exp\left(-\frac{2\omega^2 h B_{11}^2}{M_s} \cdot \frac{1}{V_{SAW}^2} \chi_{\eta\eta}'' \sin^2\phi_0 \cos^2\phi_0 |c_z|^2 z\right) \quad (22)$$

and the power attenuation of the SAW per unit width and unit length, as calculated by back-action, is given by:

$$dP_{abs}^{SAW} = -\frac{2\omega^2 h B_{11}^2}{M_s} \cdot \frac{1}{V_{SAW}^2} \chi_{\eta\eta}'' \sin^2\phi_0 \cos^2\phi_0 |c_z|^2 P_{SAW} \quad (23)$$

The magnetic oscillation power absorbed by the magnetic damping during spin-wave resonance per unit volume, is $p_{abs}^{mag} = \frac{\omega}{2} M_s \mathbf{h}_{RF}^{\dagger} \cdot \chi'' \cdot \mathbf{h}_{RF}$.[19,23] By energy conservation, the power absorbed by the magnet should equal the SAW power dissipation. The SAW power dissipation is often calculated using this so-called effective field approach. Using Eqn. (14), we express the absorbed magnetic power explicitly as

$$p_{abs}^{mag} = \omega \chi_{\eta\eta}'' \left(\frac{\beta}{\omega}\right)^2 \left(\frac{2B_{11}^2}{M_s}\right) |v_z^{(0)}|^2 \sin^2\varphi_0 \cos^2\varphi_0$$
$$= \left\{\omega^2 \chi_{\eta\eta}'' \left(\frac{1}{V_{SAW}^2}\right)\left(\frac{2B_{11}^2}{M_s}\right) |c_z|^2 \sin^2\varphi_0 \cos^2\varphi_0\right\} P_{SAW} \quad (24)$$

The spin-wave power dissipated by the magnetic damping per unit width and per unit length is $dP_{abs}^{mag} = p_{abs}^{mag} h$ and thus equals:

$$dP_{abs}^{mag} = \frac{2\omega^2 h B_{11}^2}{M_s} \cdot \frac{1}{V_{SAW}^2} \chi_{\eta\eta}'' \sin^2 \varphi_0 \cos^2 \varphi_0 |c_z|^2 P_{SAW} \qquad (25)$$

Thus $dP_{abs}^{SAW} + dP_{abs}^{mag} = 0$ as required by energy conversation (which must be satisfied at all orders of the perturbation theory). The effective field approach and the back-action approach are, in fact, one and the same. The velocity shifts arising from the spin-wave back-action are given by the real part of $\Delta\beta_{ME}^{Rayleigh}$ and is the Hilbert transform of the imaginary part of $\Delta\beta_{ME}^{Rayleigh}$. This must be the case or else causality is violated. There are, however, other field-dependent effects that can become convolved with measured wave-number shifts due to spin-wave back-action. These effects will be relevant at lower SAW pump frequencies and at low fields below the in-plane anisotropy field. In this regime there can be domain wall motion and magnetization rotation as a function of $H_{app}$. As a result, $\Delta E$ effect induced changes to the velocity[24] and Anisotropic Magnetoresistance (AMR) induced changes to the attenuation[25] will not be negligible. At higher pump frequencies and in films with low in-plane anisotropy, the magnetization will be saturated along the field direction across the spin-wave resonance field. In such cases, we expect that field dependent contributions to $\Delta\beta$ are due to $\Delta\beta_{ME}^{Rayleigh}$ and that the relation between the field-dependent part of the velocity shift and attenuation are given by Eqn. (20) at first order.

In order to get an estimate of the magnitude of these back-action effects, we calculate both real and imaginary part of $\Delta\beta_{ME}^{Rayleigh}$ vs. $H_{app}$ for a Rayleigh SAW propagating in the Z-direction on a YZ-cut LiNbO$_3$ substrate with $V_{SAW} = 3.488 \times 10^5$ cm/s and $\frac{\omega}{2\pi} = 4.5$ GHz (implying $\beta = 8.1 \times 10^4$ cm$^{-1}$). This is a regime where $\beta h = 0.08$ and perturbation solutions to first order are often reasonable. The applied field is swept at an angle $\phi_H = 45°$ with respect to the SAW propagation axis. The magnetoelastic perturbation is a Ni film with $h = 10$ nm. We have assumed that the properties of the Ni film are $H_k = 0$, $K_\perp = 5.5 \times 10^5$ ergs/cm$^3$, $M_s = 485$ emu/cm$^3$, an isotropic and wave-vector independent spin-wave damping $\Gamma = 0.1$, $B_{11} = +5 \times 10^7$ ergs/cm$^3$, and $A_{ex} = 8 \times 10^{-7}$ erg/cm. The results of the calculation are shown in Figure 3. The maximum relative shifts due to spin-wave magnetoelastic back-action are $\left|\frac{\text{Re}\,\Delta\beta}{\beta}\right| \cong \left|\frac{\Delta V_{SAW}}{V_{SAW}}\right| \sim$ .015% and $\frac{\text{Im}\,\Delta\beta}{\beta} \sim .03\%$. In a 300 μm long Ni film, this implies a phase shift from one end to the other of $\Delta\varphi \sim 25°$ and a SAW attenuation ~ -6 dB (or a power attenuation per unit length of ~ -20 dB/mm). These numbers are in accord with various experiments.[6,11,12] It is instructive to compare these wave-number shifts to those that are associated with mass loading in the Ni film. We have assumed $\rho'_{Ni} = 8.908$ g/cm$^3$, a Poisson ratio $\nu = 0.31$, and a Young's modulus $Y = 190 \times 10^{10}$ dyn/cm$^2$ and where $\frac{s'_{11}}{s'^2_{11} - s'^2_{12}} = \frac{Y}{1-\nu^2}$. Based on these values, mass loading predicts wavenumber shifts of ~0.8%. The effects on the SAW due to spin-wave back-action are thus typically an order of magnitude lower than mass-loading.

As film thickness $h$ increases, Eqn. (21) predicts that $\Delta\beta_{ME}^{Rayleigh}$ increases linearly with $h$ and depends only on the $z$ component of the particle velocity. But as the film becomes thicker, we expect that $yz$ shear strains and their impact on the stress fields within the piezoelectric will become non-negligible. Thus there will be $y$ dependent particle velocity fields at order $\mathbf{v}'^{(1)}$ in the film that can be expressed in terms of the unperturbed particle fields $v'^{(0)}_y$ and $v'^{(0)}_z$. These $\mathbf{v}'^{(1)}$ fields will generate $y$-dependent components of $\mathbf{h}_{RF}(\mathbf{r},t)$ that will then drive thickness dependent spin-wave amplitudes $\boldsymbol{\delta m}^{(1)}$. These spin-wave amplitudes $\boldsymbol{\delta m}^{(1)}$ will contribute to magnetoelastic traction forces on the SAW at second order $\mathbf{T}'^{(2)} \cdot \hat{\mathbf{y}}\big|_{y=0}$ arising from the term $B_2 m_z m_y$ in the stress tensor. Such stresses create back-action forces on the SAW that reverse sign depending on whether the projection of $\mathbf{m}_0$ onto the $z$ axis is aligned or anti-aligned with the wave-vector $\boldsymbol{\beta}$. Such effects have been observed clearly in angular dependent SAW attenuation measurements with thicker Ni films where $h = 50$ nm and where $\beta h > 0.15$.[19] Perturbation theory predicts that, at least initially, these effects must scale as $h^2$ as they result from back-action forces of second order.

We do not go through the calculation of these second order effects here. Our main point is that the perturbation theory enables one to programmatically calculate SAW attenuation and velocity shifts arising from spin-wave magnetoelastic back-action, determine at what order various effects appear, how they scale with film thickness, and what their strength is without resorting to various ad-hoc fitting parameters. Furthermore, the perturbation theory allows for a clear physical picture and realistic computational framework for how spin-wave back-action modifies time-dynamic and thickness-dependent stress/strain fields and electromechanical transmission line currents/potentials at the thin film/piezoelectric interface. Knowledge of the interfacial stress/strain fields and currents/potentials allow for extraction of various physical quantities such as the magnetoelastic coupling in the film or the magnetoelectric coupling at the magnetic/piezoelectric interface. For example, measurement of the transmission line current out of phase with the SAW potential (related by the imaginary part of $Z_{ME}^{Rayleigh}$ in Eqn. (21)) along with knowledge of $H_k$, $M_s$, and $K_\perp$ enables extraction of the magnetoelastic coupling $B_{eff}$ in a way that is separable from other phenomena affecting the transmission line impedance (e.g., mass/capacitive loading and ΔE effects).

The calculation of the various fields at the interface may also be important for a matter that we have ignored throughout the paper – the magnitude of the spin-wave damping $\Gamma$. Typical values of damping in Ni under uniform-mode ferromagnetic resonance are of order ~ 0.04.[26] However, values extracted from SAW-driven spin-wave resonance experiments are considerably larger with $\Gamma$ ~ 0.1- 0.2.[6,11,19] The spin-wave damping $\Gamma$ is, in fact, the only fitted quantity in the theory and parametrizes all the irreversible energy transfer out of the SAW/spin wave system to other degrees of freedom. It is quite plausible that this enhanced spin-wave damping $\Gamma$ is related to the back-action stress and strain fields generated by the elastically-driven spin-wave resonance (with the typical smaller magnetic damping of order $\Gamma_0^{Ni}$ ~ 0.04) and the irreversible transfer of energy out of these fields into substrate modes. Thus a perturbative calculation of the elastic back-action fields and computation of the coupling of these surface fields to bulk modes might lead to an explanation of the how the spin-wave damping $\Gamma_0^{Ni}$ becomes dressed and leads to the enhanced damping $\Gamma$ as observed in experiment. This could provide a framework for understanding dissipation of elastically-driven magnetic resonance processes and energy transfer in magnetic thin film/acoustic actuator hybrids.

**Acknowledgement:**

This work was supported in part by the NSF TANMS center. D. Labanowski gratefully acknowledges support from the NDSEG fellowship.

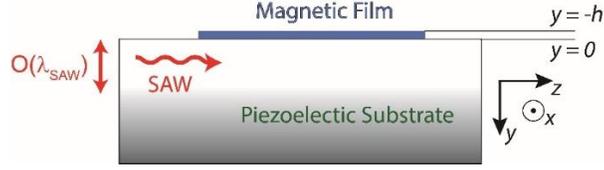

Figure 1. Coordinate system for SAW propagation and schematic of magnetic film with thickness $h$ on top of piezoelectric substrate. The film is elastically strained by a SAW traveling in the substrate with wavelength $\lambda_{SAW}$ and with a penetration depth of order $\lambda_{SAW}$ into the piezoelectric.

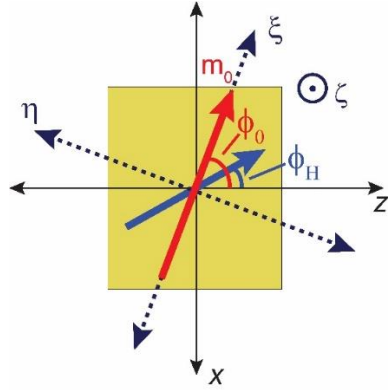

Figure 2. The $\eta\zeta\xi$ coordinate system used in LLG linearization with $+\xi$ defined to be along the equilibrium $\mathbf{m}_0$ direction. The angles $\phi_0$ and $\phi_H$ that the equilibrium magnetization and applied field make with respect to $+z$ (the SAW propagation direction) have also been defined.

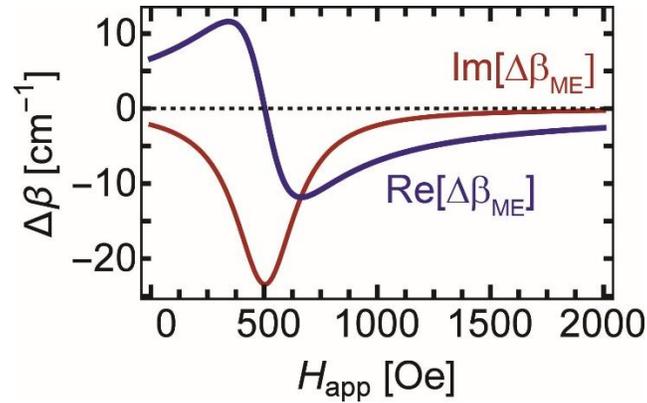

Figure 3. Calculated shifts in the wave-number $\Delta\beta_{ME}^{Rayleigh}$ of a Z-propagating Rayleigh SAW on YZ-cut LiNbO$_3$ vs. $H_{app}$ due to the magnetoelastic backaction of a spin-wave resonance in a 10 nm thick Ni film. The field is swept at $\phi_H = 45°$ with respect to the Z axis.

22